\documentclass[sigconf]{acmart}
\AtBeginDocument{%
  }

\acmISBN{978-1-4503-XXXX-X/2026/06}
\usepackage{tabularx,booktabs}
\usepackage[title]{appendix}
\usepackage{CJKutf8}
\usepackage{subcaption}
\usepackage{tabularx}      
\usepackage[table]{xcolor} 
\usepackage{makecell}
\usepackage{enumitem}
\usepackage{longtable}
\usepackage{booktabs}
\usepackage{multirow}
\usepackage{listings}
\usepackage{fancyvrb}
\usepackage{array}
\usepackage{colortbl}
\usepackage{ragged2e}
\usepackage{adjustbox} 
\usepackage[normalem]{ulem}
\usepackage{pgfplots}

\newcolumntype{L}[1]{>{\RaggedRight\arraybackslash\hsize=#1\hsize}X}

\newcommand{\BlockHeightTaller}{\rule{0pt}{15pt}}  
\usepackage[table]{xcolor}   
\usepackage{collcell}        

\definecolor{ColorBaseline}{HTML}{BFDCCC}
\definecolor{ColorPrompt}{HTML}{FFD2C2}
\definecolor{ColorAgent}{HTML}{C5D1F5}

\newcommand{\ShadeParticipants}[1]{%
  \if\relax\detokenize{#1}\relax 
    \cellcolor{white}%
  \else
    \ifnum#1=0 \cellcolor{white!100}\fi
    \ifnum#1=1 \cellcolor{white!85}\fi
    \ifnum#1=2 \cellcolor{white!70}\fi
    \ifnum#1=3 \cellcolor{white!55}\fi
    \ifnum#1=4 \cellcolor{white!40}\fi
    \ifnum#1=5 \cellcolor{white!25}\fi
    \ifnum#1=6 \cellcolor{white!10}\fi
    \ifnum#1=7 \cellcolor{white!0}\fi
  \fi
}

\newcolumntype{S}[1]{>{\collectcell{\ShadeParticipants}}m{#1}<{\endcollectcell}}

\begin{document}

\title{Enacting Constructive Conflicts with AI Agents to Enhance Reconsideration among Novice Interaction Designers}



\renewcommand{\shortauthors}{Han et al.}

\author{Howard Ziyu Han}
\orcid{0009-0008-5556-7297}
\email{ziyuh@andrew.cmu.edu}
\affiliation{%
  \institution{Carnegie Mellon University; Human-Computer Interaction Institute}
  \city{Pittsburgh}
  \state{Pennsylvania}
  \country{USA}
}

\author{Nikolas Martelaro}
\orcid{0000-0002-1824-0243}
\email{nikmart@cmu.edu}
\affiliation{%
  \institution{Carnegie Mellon University; Human-Computer Interaction Institute}
  \city{Pittsburgh}
  \state{Pennsylvania}
  \country{USA}
}


\begin{abstract}
Generative AI agents are increasingly used in interaction design to facilitate ideation and offer critique, often following their own internal reasoning. These interactions tend to add design ideas and expand the design space. Our work explores an antagonistic role for design agents, prompting designers to engage with stakeholder tension. We built an AI agent inspired by adversarial design theory that enacts constructive conflict. We examine the agent's influence in a between-subjects experiment with 45 design students across three conditions: Self Reflection (unsupported review of the design proposal), Stepwise Guidance (written prompts that walk designers through a constructive-conflict framework), and Interactive Engagement (an AI agent that enacts the constructive-conflict framework interactively by synthesizing stakeholder pushback). The latter two conditions share the framework but differ in whether it is self-enacted or agent-enacted. Results show that, compared with Self Reflection, both the Stepwise Guidance and Interactive Engagement groups reported significantly higher self-reconsideration and made more improvements to their design proposals. Compared with Stepwise Guidance, the antagonistic agent introduced more conflictual perspectives, and participants in the Interactive Engagement condition generated and discarded more ideas. These findings suggest that agent-enacted constructive conflict can turn reconsideration into concrete design actions and deepen engagement with divergent stakeholder perspectives.
\end{abstract}
\setcopyright{none}                           
\settopmatter{printacmref=false}               
\renewcommand\footnotetextcopyrightpermission[1]{}     
\pagestyle{plain}                              
\maketitle

\section{Introduction}
Computational agents are increasingly used in interaction design, shaping how designers explore possibilities and develop design outcomes~\cite{long2025feedquac, wang2025aideation, chiou2023designing, naka2025creative}. In doing so, these agents do not simply support the creation of design artifacts. They introduce design perspectives and steer trajectories through their own forms of agency, which may differ from designers' habitual modes of thinking. Such influence becomes particularly consequential in multi-stakeholder interaction design---for example, public-sector design, where civic platforms must serve residents, agencies, and advocacy groups whose interests often conflict. These are classic wicked problems~\cite{buchanan1992wicked}: competing values make them difficult to define, let alone optimize for. 
In these contexts, using AI agents can affect how stakeholder perspectives are introduced and interpreted, thereby shaping how stakeholder tensions are translated into design decisions.

One framework to approach this design mediation of multi-stakeholder issues is \textit{constructive conflict}. Rather than resolving differences, approaches such as adversarial design treat conflict as a productive mechanism for surfacing assumptions and expanding the design space~\cite{muller_participatory_1993, geppert2022design, disalvo2010design}. Drawing on agonistic pluralism~\cite{mouffe1999deliberative}, adversarial design frames design as an ongoing negotiation in which competing viewpoints are sustained. This opens a different line of inquiry for Human-Agent Interaction (HAI) : not how agents resolve conflict, but how they keep it alive.

In HAI research, large language model (LLM)-powered devil's advocate agents have been shown to improve decision-making by challenging human teams~\cite{chiang2024enhancing}. Design contexts, however, are marked by asymmetrical collaboration and compound decisions. Thus, devil's advocate agents designed for decision support may not directly apply to design, where conflict can shift the problem space rather than only improve judgment within it. 

In this paper, we investigate two research questions:

\noindent\textbf{RQ1: How can an interactive AI agent be thoughtfully designed to enact constructive conflict within multi-stakeholder interaction design contexts, such as public-sector design?} We approach this question by learning from six experienced public-sector designers and prototyping an LLM agent that helps designers engage with stakeholder pushback during early design ideation.

\noindent\textbf{RQ2: How might AI agent-mediated conflict shape novice designers' approaches to constructive conflict in design contexts with multiple stakeholder considerations?} We engaged 45 interaction design students to redesign a local civic reporting website. We focus on design students because they are trained in interaction design but novices at navigating stakeholder conflict. Additionally, novice designers increasingly use LLMs that often have a tendency toward agreement.
This suggests that exploring anti-sycophantic interaction could be especially consequential for how designers work in the future~\cite{cheng2026sycophantic}. Participants were assigned to three conditions: reflection on their design work without structured guidance (Self Reflection), stepwise prompts on thinking about multi-stakeholder conflict (Stepwise Guidance), and interacting with an agent (Interactive Engagement). Notably, the stepwise prompts in the Stepwise Guidance condition mirror the Interactive Engagement agent's system prompts, allowing us to disentangle the contributions of the underlying thinking framework from the interactive process of engaging with an agent. 

We find that interaction design students who engaged with the agent reported significantly more reconsideration of their design decisions, more shifts in design thinking, and more iterative edits to their design proposals compared to Self Reflection. The Stepwise Guidance condition also encouraged reconsideration but did not shift overall thinking, and even reduced new idea generation. Together, these findings suggest that stakeholder tension---whether conceptually introduced or enacted through an AI agent---deepens reconsideration, with the agent offering a more balanced mix of reflection and generativity. Such differences between stepwise guidance on stakeholder tensions and the direct enactment of those tensions through the agent point to the potential for interactive AI systems to scaffold more active engagement with stakeholder conflict, supporting more robust design processes.

\section{Related Work}

\subsection{AI Agents for Design Feedback and Thinking about Stakeholders}
Research in Human-AI Interaction has highlighted the potential of AI agents to support design work such as offering suggestions~\cite{liang_encouraging_2024,shin_introbot_2023,ataei_elicitron_2024}, augmenting cognitive capacities~\cite{ariely2011thinking}, encouraging critical reflection~\cite{orchard2024fostering}, and counteracting bias~\cite{design2006we}. This lineage traces back to Fischer's early critic agents, which surfaced ``breakdown situations'' in designers' reasoning to prompt deeper understanding and creative problem-solving~\cite{fischer1993embedding}.
More recent LLM-based agents employ techniques such as open-ended dialogue, persona simulation, and alternative framings to support design critique~\cite{liuPersonaFlowDesigningLLMSimulated2025, ataei_elicitron_2024}.

Some of these agents are argued to be valuable for novice designers, exposing them to expert reasoning and industry guidelines while functioning as on-demand tutors or mentors~\cite{duan2024uicrit,karwa2023ai}.
Across these efforts, however, agents are predominantly tuned for helpfulness and agreement. A recent thread of HAI work has begun to introduce friction through metacognitive prompting, interrupting habitual reasoning with reflective questions~\cite{gmeiner2025exploring} or cognitive forcing functions to slow analytical engagement~\cite{buccinca2021trust}.
Yet these interventions ask the user to think harder while the agent itself holds no contrary stance. We extend this thread by investigating antagonistic agents that go beyond metacognitive nudging to genuinely push back, positioning the designer as a deliberator engaged in substantive disagreement rather than a recipient of suggestions.

\subsection{Constructive Conflicts in Design Processes}
We ground our approach to constructive conflict in agonistic pluralism, a framework that design scholars have adopted to treat stakeholder disagreement as productive rather than something to be resolved~\cite{disalvo2010design, bjorgvinsson2012agonistic}. Originating in political theory, agonistic pluralism reframes collective decision-making as ``\textit{grounded in productive conflict},'' where competing values are continuously contested rather than reconciled~\cite{mouffe1999deliberative}. Building on this stance, adversarial design creates ``\textit{spaces of confrontation}'' to challenge assumptions and practices~\cite{disalvo2010design}, and social justice-oriented design frameworks treat a \textit{commitment to conflict} as foundational~\cite{dombrowski_social_2016}. Together, these lines of work position \textbf{constructive conflict} as a means to reveal hidden assumptions and negotiate design implications.

Operationalizing constructive conflict is easier said than done. In participatory design practice, there is often a push toward harmony and consensus, which can inadvertently silence important disagreements~\cite{gautam2024surfacing,bjorgvinsson2012agonistic}.
Even when a project aspires to procedural justice (i.e., giving all stakeholders a voice), simply inviting tensions into the room does not automatically translate into better design outcomes, and can even fall into tokenism~\cite{dalsgaardParticipatoryDesignLargeScale2012}. The gap between the conceptual productivity of conflict and the practical challenge of surfacing it highlights the need for approaches that help designers actively engage with stakeholder tensions.

\subsection{Devil's Advocate Agent in Teams for Improved Decision Making}
Management science has long noted that a certain level of team conflict can enhance group outcomes by preventing homogeneous groupthink~\cite{schweiger1986group,chen2006understanding}. Techniques like the Devil's advocate, i.e., assigning someone to intentionally challenge the prevailing opinion, can lead to more creative, higher-quality decisions~\cite{parkerAntagonismAccommodationAgonism2017,markhamChampionsAntagonistsRelationships1991}. The key insight is that exposing ideas to critique yields a more thorough analysis than uncritical consensus~\cite{schweiger1986group,herbert1977improving}. However, having a human team member constantly playing the antagonist role can be difficult both operationally and interpersonally~\cite{jamieson_sympathy_2014,nemeth2001devil}. Research shows that individuals tasked as a devil's advocate can struggle to argue points they do not genuinely believe, leading to less persuasive or superficial counterarguments~\cite{nemeth2001devil}. In addition, the person in this role can face social repercussions, such as being disliked or feeling isolated for continually opposing others~\cite{jamieson_sympathy_2014}.

AI agents may offer a compelling alternative to team members as a devil's advocate. The AI agonist can perceive the discussion context, plan when and how to interject and deliver critical questions or counterpoints~\cite{cai2024antagonistic}. By offloading the dissenting role to a nonhuman entity, teams might also reap the benefits of agonistic debate for more contested decisions~\cite{kripleanSupportingReflectivePublic2012a,chen2025maintaining}, and individuals might learn how to resolve conflicts~\cite{shaikhRehearsalSimulatingConflict2024}. Chiang et al. introduced an LLM-powered devil's advocate into risk-prediction tasks with non-design teams and found that it could improve decision outcomes under the right conditions~\cite{chiang2024enhancing}. The agent challenged the group's conclusions, leading teams to scrutinize information more carefully without derailing the decision process.

However, this body of work concentrates on convergent decision tasks with verifiable answers and non-design teams. Adversarial agents remain largely unstudied in design contexts, where decisions lack a single correct solution, and the problem space itself can shift. We bridge this gap by extending the devil's-advocate paradigm into interaction design, asking how disagreement should unfold turn-by-turn between agent and designer and how sustained antagonistic exchange shapes designers' reasoning and final proposals.

\section{Methods}
\subsection{Agent Design and Workflow}
\begin{figure*}
    \centering
    \includegraphics[width=\linewidth]{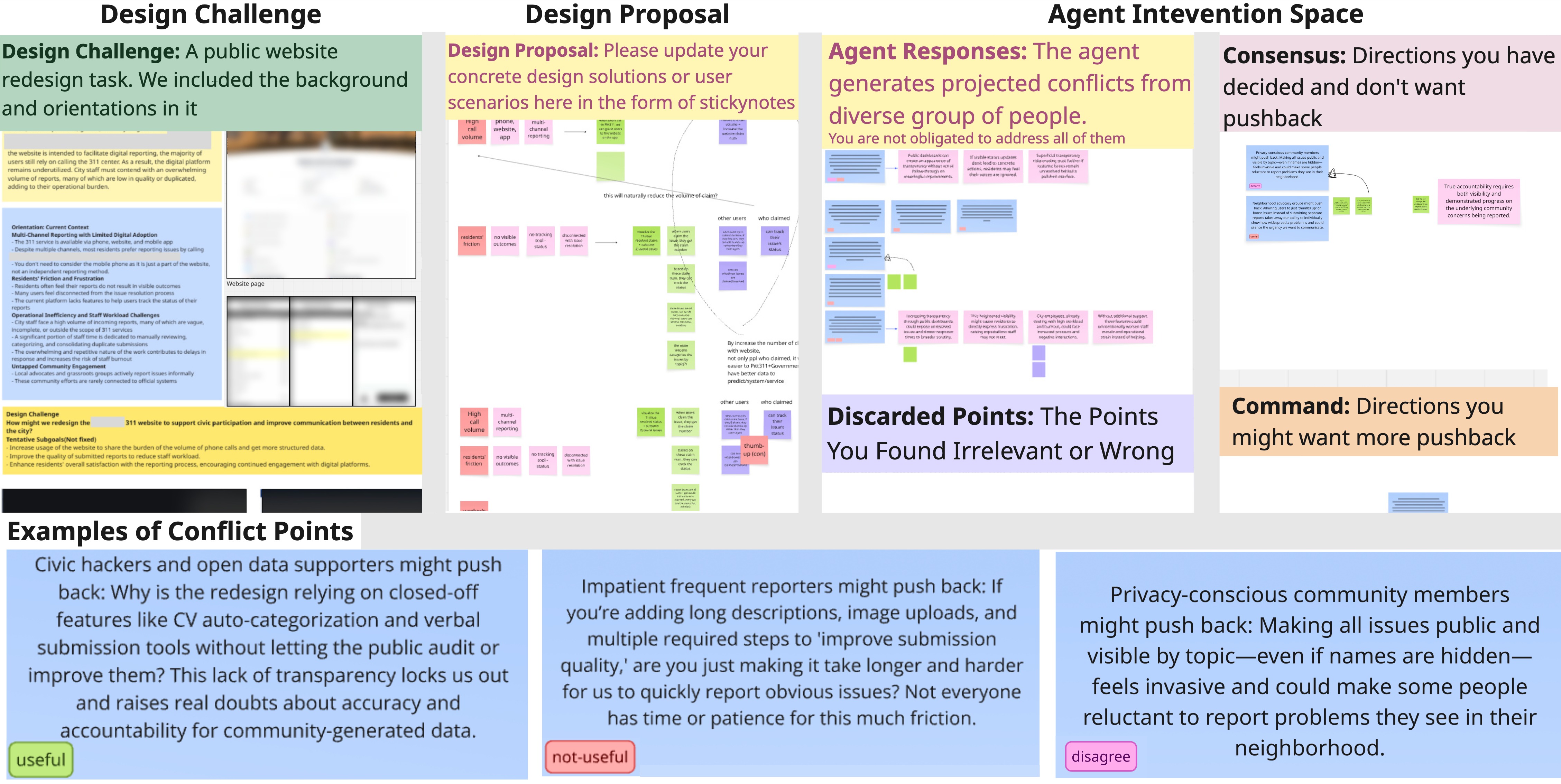}
    \caption{The \textbf{Interactive Engagement} condition's Miro board, showing (left to right) the design challenge brief, the participant's design proposal space, the agent interaction space with \textit{Consensus} and \textit{Command} frames for steering pushback, and example conflict points in a Stakeholder+Pushback format. Participants tagged each point as useful, not useful, or with custom tags.}
    \label{fig:system}
\end{figure*}

The antagonistic agent's design was grounded in formative interviews with six professional designers experienced in multi-stakeholder and conflict-sensitive public design work ($M=13.5$, $SD=9.20$ Years of experience, demographics in Appendix~\ref{appen:experts-participants}). \textbf{Four priorities emerged consistently, which led us to develop four design goals (DG) for our agent.} First, constructive conflict should be stakeholder-grounded and evolve as the design progresses, rather than remaining static and generic (DG1). Second, designers wanted to retain agency while developing their ideas, engaging with critique structurally rather than as a stream of suggestions, which distinguishes our approach from conversational devil's-advocate agents~\cite{chiang2024enhancing} (DG2). Third, constructive conflict should stress-test directions designers are committing to and help them reframe design goals entirely (DG3). Finally, since designers cannot resolve every conflict, they must learn to steer through it. The professionals suggested that a system should make trade-offs explicit by naming which values to prioritize and which to defer, rather than forcing false resolution across all conflicts (DG4).

Built upon these goals, we implemented the agent in Miro\footnote{https://miro.com/}, a digital whiteboard. The workspace comprises three areas: 1) Design Challenge, 2) Ideation Board, and 3) Agent Intervention Space (see Fig.~\ref{fig:system}), keeping unprompted ideation distinct from agent critique (DG2). As designers work, the agent passively gathers multimodal context (text, visuals, and participants' thinking-aloud), so that its pushback is anchored in designers' contextual exploration (DG1). The professional designers also stressed that constructive conflict needs bite, so we deliberately designed the agent's tone to be adversarial rather than polite (see system prompt, Appendix~\ref{app:system-prompt}).

Within the Agent Intervention Space, the activity unfolds as follows: the agent generates a first round of four pushback points anchored in the designer's current proposal; the designer interacts with those points to steer the design direction; the agent then generates a second round informed by that steering. Each \textit{pushback point} is a sticky note pairing a stakeholder perspective with a specific challenge to the designer's proposal (DG1). The first round is open-ended: the agent surfaces tensions it judges most relevant to the designer's current proposal, exposing designers to perspectives they may not have anticipated (DG1). Before the second round, the designer steers the agent through two mechanisms: (1) \textbf{point-level tagging}, in which they label each pushback point as useful, not useful, or with a custom tag, and discard irrelevant ones; and (2) \textbf{stance-level framing}, in which two optional frames declare the designer's position, namely a Consensus frame for decisions to consolidate and a Command frame for areas inviting additional pressure. The second round then narrows to the directions designers chose to probe further. Together, these mechanisms let designers stabilize emerging directions while deliberately exposing them to further critique (DG3) and articulate the trade-offs (DG4).

To ground these critiques in real practice, we compiled a database of prior constructive conflict examples shared by the professional designers, alongside documents on agonistic pluralism and constructive conflict. A sub-agent retrieves and synthesizes these materials when generating each round of pushback points.

\subsection{Agent Preliminary Evaluation and Implementation}
Before deployment, we iteratively refined prompts and ran three pilot tests with design students recruited separately from the formative-interview experts and from the main user study participants. Each pilot participant completed the design challenge themselves and then rated ten critique points, five from Gemini 2.5 Pro and five from GPT-4.1, on four dimensions following prior HAI evaluation work~\cite{yen2024give,mukherjee_impactbot_2023}: introducing friction from stakeholders, helping the designer think about diverse perspectives, projecting stakeholders' perspectives reasonably, and contributing to design thinking. Both models produced comparable outputs that participants judged reasonable. GPT-4.1 received ratings between 5.00 and 5.67 (SDs 0.58--2.08), and Gemini 2.5 Pro received means between 5.33 and 5.67 (SDs 0.58--1.15) on a 7-point Likert scale. The two models tied on two of the four items, with no model consistently outperforming the other. We thus selected GPT-4.1 for the user study because its faster generation maintained a productive design cadence.

The agent was built with Next.js, the Miro SDK, and the Miro REST API. GPT-4.1 generates critique points, Firestore stores interaction data, Whisper processes voice input, and GPT-4o parses visual content. We reused the Miro's native UIs such as stickynotes. A few actions not yet exposed by Miro's open API\footnote{https://developers.miro.com/} (e.g., buttons to generate pushback points) were handled by the researcher via a predefined, bias-avoiding protocol (see Appendix~\ref{app:WoZ}). The researcher sent points only via fixed triggers and did not elaborate on the agent's points. The codebase will be released upon acceptance.

\subsection{Interaction Design User Study}
\label{sec:user-study}
To understand how the antagonistic agent influences the interaction design proposal, the iteration process, and the designers' thinking when working on a multi-stakeholder design challenge, we conducted a between-subjects study with interaction design students across three conditions: 1) Self Reflection, 2) Stepwise Guidance, and 3) Interactive Engagement.

\subsubsection{Task}
We chose the redesign of a local government non-emergency civic reporting website, commonly known in the US as a ``311'' site\footnote{see \url{https://en.wikipedia.org/wiki/3-1-1}. Many other countries operate equivalent services, such as local council reporting portals in the UK and the 12345 citizen hotline in China.} as the task. On such a site, residents submit \textit{tickets} to report non-emergency civic issues (e.g., potholes, graffiti, illegal dumping). City staff then route, triage, and resolve these tickets. This task involves interaction design for a mix of stakeholders, including city staff, diverse residents, and community organizations. Participants were asked to complete an interaction design proposal focused on redesigning the website to achieve the design goals across two iterations. The full design brief is in Appendix~\ref{app:design-challenge}.

\subsubsection{Participants and Apparatus}
We recruited 45 design students from Design and Information Science departments, with each randomly assigned to one of the three groups. Their average age was 25.08 ($SD = 3.68$), with an average of 4 years of studying design/HCI ($SD = 1.79$). Self-reported familiarity with interaction design was high ($M=5.8$, $SD=1.3$, on a 7-point scale, 7 = ``very familiar''). The study was conducted online via Zoom and lasted approximately 90 minutes. Each participant received a \$30 compensation. The study was approved by our university IRB.


For the user study, the procedures are as follows:
\begin{itemize}[left=0pt]
  \item \textbf{Introduction (10 minutes):} The participants were introduced to using Miro and asked to review the design brief for updating a local non-emergency reporting website. Participants assigned to the Interactive Engagement condition were additionally informed that the critique points they would later see were synthetic and generated by an AI agent.
  \item \textbf{Design Iteration 1 (20--30 minutes):} The participants worked for 20--30 minutes on their initial design ideation, jotting down ideas in the Miro canvas. In this stage, all three groups worked without any interventions. For the Interactive Engagement group, the agent passively collects context during this stage.
  \item \textbf{Condition Phase (15 minutes\footnote{The time was determined based on the pilot tester's time spent interacting with the agent and comprehending the reflective awareness material.}):} After completing the first iteration, participants were assigned to one of the three conditions.
  
\begin{itemize}[left=0pt]
    \item \textbf{Self Reflection:} The Self Reflection group was asked to review and reflect on their design proposal, then think aloud about what they had reflected on. The reflection prompt given to participants is in Appendix~\ref{app:prompts}.
    \item \textbf{Stepwise Guidance:} The Stepwise Guidance group reviewed a stepwise introduction to enacting constructive conflict, with steps adapted from the system prompt used by the agent (full prompts and their correspondence in Appendix~\ref{app:prompts}). Participants read the material and thought aloud.
    \item \textbf{Interactive Engagement:} The Interactive Engagement group first saw four pushback points generated by the agent, interacted with the agent to steer the next round, and then saw a second set of four points. We chose four based on pilot tests, balancing breadth of perspectives against cognitive load.
\end{itemize}

  \item \textbf{Design Iteration 2 (15--25 minutes):} After the intervention, participants were asked to do another iteration by adding, editing, or deleting any ideas from the first iteration.
  \item \textbf{Post-Survey \& Interview (15 minutes):} All participants completed a post-survey on their perceived design outcome and design thinking. The Interactive Engagement group filled out an additional short survey on their experience with the AI agent. All participants were then interviewed about their overall experience and thoughts on the iterations.
\end{itemize}

\subsubsection{Measurements}
\label{sec:measure}
To understand how design behaviors changed across groups, we measured the number of design ideas generated across the two iterations using three methods: (1) we identified each \textbf{idea unit}, defined as one or more sticky notes that together represent a single design proposal element; (2) we used participants' continuous think-aloud during ideation to disambiguate newly added ideas from edits to existing ones; and (3) we used the verbal walkthrough participants gave at the end of each iteration to verify unique ideas across coders. Each idea was labeled as added, deleted, or edited between the two iterations, with edits split into \emph{revisions} (partial deletion and restart) and \emph{enhancements} (redevelopment of an existing idea). Enhancements were distinguished from additions via proposal cues (enhancement notes connect to existing ideas and are incomplete) and think-aloud phrasing (e.g., ``we can also add''). Ambiguous cases were resolved by team discussion; the full codebook is in the supplementary material.

All participants completed a post-survey on 7-point Likert scales measuring perceived challenge and satisfaction~\cite{neeley2013building,vignoli2023design}, self-efficacy~\cite{chen2001validation,dow2010parallel}, the amount of design change between iterations~\cite{wynn2017perspectives}, and critical design thinking~\cite{vignoli2023design,redhana2020validity}, with optional qualitative follow-ups. The Interactive Engagement group additionally rated their experience working with the AI agent. This included collaboration experience~\cite{borsciChatbotUsabilityScale2022}, results-worth-effort, enjoyment~\cite{cherry2014quantifying}, agency, and the agent's perceived effectiveness on four design considerations~\cite{oh2018lead,yen2024give} (the four considerations and the full survey are in Appendix~\ref{app:survey}). For the Interactive Engagement condition, we also counted the tags participants applied to pushback points and analyzed their steering activity on the Miro board.

\subsection{Data Analysis}
From the study, we collected 45 design proposals, 90 iterations, and approximately 60 hours of transcripts.

\subsubsection{Qualitative coding} The first author coded all design proposals to extract the idea units and behaviors of added, edited, and deleted ideas, following the methods in the measurements and prior research~\cite{xu2025productive}. The proposals and transcripts were analyzed through mixed deductive--inductive qualitative coding, with codes sorted into affinity diagrams and refined through team consensus until broader patterns emerged~\cite{fereday2006demonstrating}.

\subsubsection{Quantitative analysis}
For the post-survey data and quantitative design behavior change, we assessed data normality using Shapiro-Wilk tests, and homogeneity of variance was evaluated using Levene's test. Kruskal--Wallis was used for non-normally distributed data, and one-way ANOVA for normally distributed measures. Post-hoc pairwise comparisons were conducted using Tukey HSD for ANOVA and Dunn's test with Bonferroni correction for Kruskal--Wallis to control family-wise error rates across three pairwise comparisons. We also report effect sizes alongside every inferential test: epsilon-squared ($\varepsilon^2$) for Kruskal--Wallis and omega-squared ($\omega^2$) for ANOVA as omnibus effect sizes, and Hedges' $g$ (Cohen's $d$ with a small-sample correction) for pairwise contrasts.

%
%

%
%

\section{Findings}
We organize our findings around RQ2, reporting our quantitative measures
alongside context from our qualitative data on how the agent shaped design behaviors, thinking, and how participants used and perceived the agent.

\subsection{How does the antagonistic agent influence design proposals?}
\begin{figure}[h]
    \centering
    \includegraphics[width=0.9\linewidth]{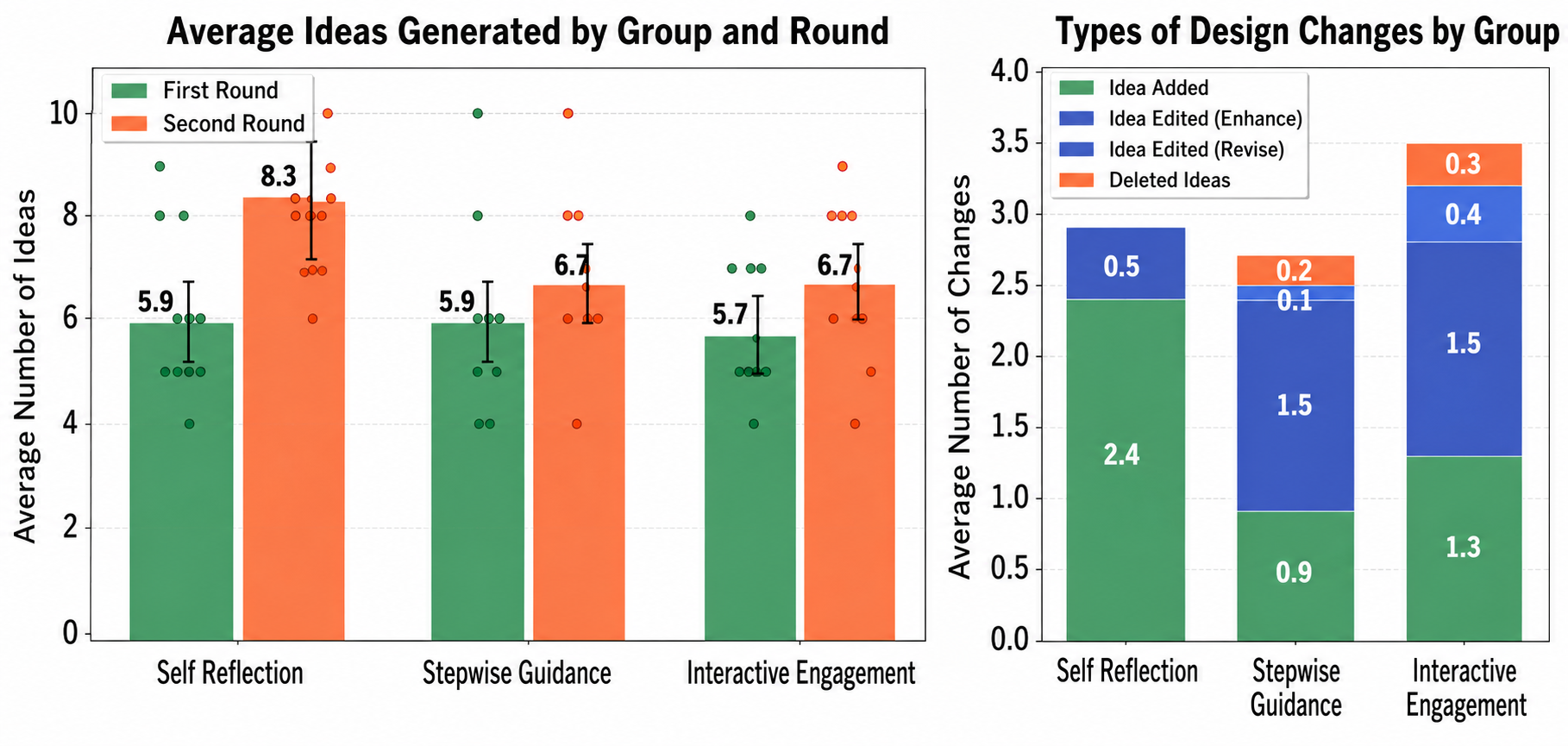}
    \caption{The left panel shows the
    average number of ideas generated in two rounds. The right
    panel shows the average number and types of design changes. 
    Error bars indicate 95\% confidence intervals.}
    \label{fig:behavioral-overview}
\end{figure}
In the first round, three groups generated similar numbers of ideas
(Self Reflection: $M=5.93$, $SD=1.39$; Stepwise Guidance: $M=5.93$, $SD=1.58$; Interactive Engagement: $M=5.67$, $SD=1.35$).
After the iteration, the Self Reflection group had the most ideas in the second round
($M=8.33$, $SD=2.06$; see Fig.~\ref{fig:behavioral-overview}).
The Interactive Engagement group \textit{changed} the most ideas overall ($M=3.47$, $SD=1.51$),
but this difference was not significant ($H(2)=1.20$, $p=.548$).

\subsubsection{Idea Addition}
\begin{figure}[h]
    \centering
    \includegraphics[width=\linewidth]{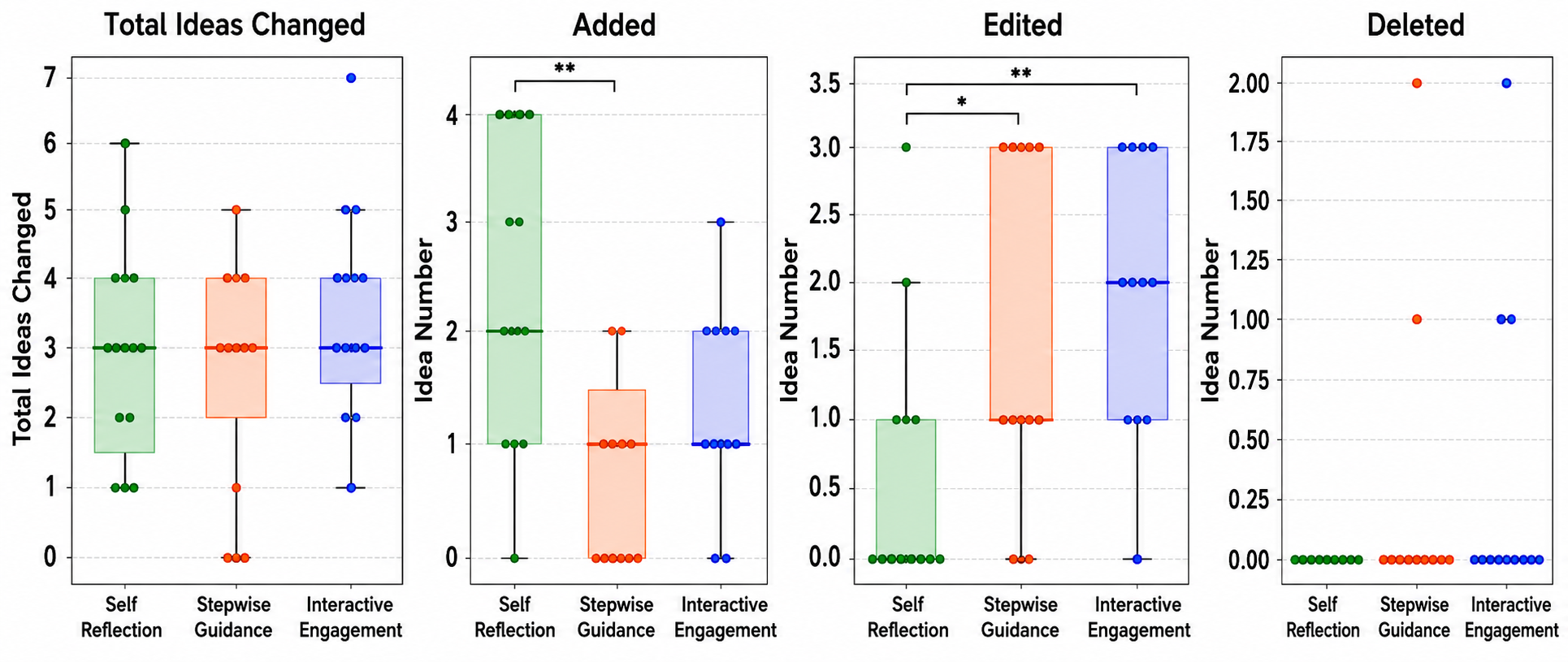}
    \caption{The figure shows how ideas changed between iterations through
    additions, edits, and deletions. *Stars indicate statistically significant
    differences ($*p < .05, **p < .01$).}
    \label{fig:behavioral-change}
\end{figure}

We define \textit{idea addition} as participants adding ideas in the second
iteration that are different from the ideas in their first iteration. The results
suggest that thinking about stakeholder tension alone might slow down designers
from producing new ideas.
For design ideas added during the second round, a Kruskal-Wallis test indicated
significant group differences ($H(2)=9.12$, $p=.010$, $\varepsilon^2=.21$, large effect; see Fig.~\ref{fig:behavioral-change}).
Post-hoc tests showed that participants in the Stepwise Guidance condition ($M=0.80$, $SD=1.15$)
added significantly fewer ideas than those in the Self Reflection condition ($M=2.13$, $SD=1.36$;
$p_{adj}=.008$, Hedges' $g=-1.03$, large effect). The Interactive Engagement condition ($M=1.27$, $SD=0.80$)
fell between the two. 
Neither contrast was significant after Bonferroni correction (vs.\ Self Reflection, $p_{adj}=.446$; vs.\ Stepwise Guidance, $p_{adj}=.346$).

Notably, three participants in the Stepwise Guidance group (P4, 11, 18-Guidance)
did not generate any new ideas or revise existing ones. P4-Guidance explained: ``\textit{I would
need to spend a day or two on this to know exactly what the tension looks like...}.''
Even participants who added new ideas expressed hesitation. For example, P28-Guidance commented:
``\textit{Having the stakeholders' questions in mind... I also questioned
whether the stakeholders would approve of the new ideas. Then that question brought all those ideas down.}''

In contrast, the \textbf{agent appeared to play a more balanced role in helping
designers transform stakeholder tensions into design ideas}, partly because of the
concrete and specific counterarguments it provided.
Rather than generating ideas in the abstract, participants P2-Engagement,
P26-Engagement and P37-Engagement added new ideas in direct response to specific
pushback points the agent raised. For example, P12-Engagement added two ideas in response to the agent's point that city staff might face additional errors
introduced by AI-augmented filters.

\subsubsection{Idea Edit}
We define an \textit{idea edit} as either a revision (rewriting an existing idea)
or an enhancement (adding more detail to it).
A Kruskal--Wallis test indicated significant group differences ($H(2)=12.42$, $p=.002$,
$\varepsilon^2=.28$, large effect). Dunn's post-hoc tests
with Bonferroni correction showed that both the Interactive Engagement condition ($M=1.93$, $SD=0.96$;
$p_{adj}=.002$, Hedges' $g=1.45$, large effect) and the Stepwise Guidance condition ($M=1.60$, $SD=1.24$;
$p_{adj}=.031$, $g=0.95$, large effect) exhibited substantially more editing than the Self Reflection
($M=0.53$, $SD=0.92$). 
The comparison between Stepwise Guidance and Interactive Engagement was not significant.
The magnitude of these effects suggests not merely a marginal increase but a shift
in iteration behavior. Interactive Engagement participants edited roughly 3.6$\times$ more ideas than Self Reflection group on average.

Enhancements were significantly higher in both the Stepwise Guidance ($p_{adj}=.036$) and Interactive Engagement
($p_{adj}=.030$) than in the Self Reflection, with no difference between Guidance and Engagement.
Revisions were significantly higher in the Interactive Engagement group than in the Self Reflection ($p_{adj}=.023$),
with no difference between Stepwise Guidance and Self Reflection.
Together, these results suggest that the Stepwise Guidance group primarily engaged in enhancements, whereas the Interactive Engagement group supported both revisions and enhancements, potentially contributing to a more reflective iteration process.

Five participants in the Interactive Engagement group revised a total of six ideas. Surprisingly,
participants in the Self Reflection group \textbf{\textit{did not}} revise any ideas, and only one
participant in the Stepwise Guidance group revised one idea. For example, P28-Guidance proposed a new
feature of showing all tickets for navigation, then revised the interaction flow to better align
with other features. In the Interactive Engagement group, participants changed mechanisms in direct response
to the agent's pushback.
For instance, the agent raised the concern that community members might feel their efforts
were devalued when tickets are merged. In response P14-Engagement revised the idea that
``algorithms automatically merge the tickets'' into ``add [users'] name to an existing ticket as
secondary if it is already in the system.'' Similarly, the agent raised the point that senior
citizens might not use smartphones or computers; in response, P37-Engagement revised the idea of 
``\textit{chatbot resolves any questions online}'' into ``\textit{chatbot records information and redirects the
conversation to the city call center}.''
\subsubsection{Idea Deletion}
Deleting ideas was a less frequent behavior across participants
(see Fig.~\ref{fig:behavioral-change}). Three participants in the Interactive Engagement group
(P19, 21, 34-Engagement) deleted a total of four ideas, while two
participants in the Stepwise Guidance group (P15-Guidance, P42-Guidance) deleted three ideas.
\textbf{\textit{No deletions}} occurred in the Self Reflection group, even though participants were
explicitly told that deletion was allowed.
Deletions appeared to occur only when participants realized friction from potential disagreement.
For example, P34-Engagement valued the agent's pushback about equity concerns in their
gamification idea. The agent raised a conflicting voice that if more engaged users gained
disproportionate attention, the system might turn into a popularity contest. P34-Engagement
reflected: ``\textit{Oh, the question is almost about seeing this platform as a democratic system.}''
They then deleted two features related to prioritizing requests from more active users.

\subsection{How does the agent influence designers' stakeholder reasoning and self-assessment?}
\subsubsection{Stakeholders considered.}
During the intervention, participants in the Self Reflection group most often reflected on and analyzed
their current design proposals (9 participants), followed by
revisiting the design brief or the existing website (8 participants).
In terms of stakeholders considered, the Self Reflection group surfaced relatively few: five
participants mentioned thinking about different stakeholders, primarily city staff
(P44-Reflection, P16-Reflection) and residents who might prefer alternative reporting channels
(P9-Reflection, P39-Reflection, P5-Reflection). The Stepwise Guidance group considered a wider range,
most frequently accessibility (7/15) and privacy (7/15). For the Interactive Engagement group, we analyzed
the stakeholder points generated by the agent, which consistently introduced accessibility,
privacy, reluctance toward automation, and preferences for alternative reporting channels.

Comparing across conditions
 the pattern is
that all stakeholder topics raised in the Self Reflection group, and most of those raised in the
Stepwise Guidance group, were also surfaced by the agent in the Interactive Engagement condition, which additionally
introduced topics neither group reached on its own. This suggests that the agent can
function as an interactive checklist for common stakeholder concerns beyond what a single human designer could think of.

\subsubsection{Goal achievement and serving diverse users.}
The Interactive Engagement group reported the highest self-assessed ratings of achieving the design goal
($M=5.40$, $SD=0.74$), compared with Self Reflection ($M=4.87$, $SD=1.30$) and Stepwise Guidance
($M=5.07$, $SD=0.80$). For serving diverse user groups, the Interactive Engagement group again reported the
highest values ($M=4.80$, $SD=1.32$) compared with Stepwise Guidance ($M=4.40$, $SD=1.30$) and Self Reflection
($M=4.13$, $SD=1.25$). None of these omnibus differences were significant. 
The more revealing pattern is a gap between self-perception and behavior. Although Stepwise Guidance and Interactive Engagement participants rated their own designs no higher, and often lower on diverse-user items, they actually considered a broader set of stakeholders than the Self Reflection group. Being prompted to consider diverse users seems to have made them aware that the range was wider and harder to address than assumed, leading them to rate their designs more conservatively even as their scope expanded.

\subsubsection{Self-efficacy and design iteration.}
Self-efficacy ratings showed no significant differences across groups. A pattern emerged, however,
with the Stepwise Guidance group generally less confident than the Self Reflection, while the Interactive Engagement group
reported the highest self-assessments. For example, on confidence in future public design, the
Interactive Engagement group ($M=5.60$, $SD=1.12$) reported higher scores than the Stepwise Guidance group
($M=5.00$, $SD=1.20$; $g=0.50$, medium effect), with the Self Reflection group in between
($M=5.53$, $SD=1.06$). Although these differences did not reach significance, the medium-sized 
effect between Interactive Engagement and Stepwise Guidance suggests a potentially meaningful divergence in confidence
trajectories that warrants replication with a larger sample. Several Interactive Engagement participants noted
feeling more confident because their ideas had been ``tested'' against antagonistic AI input
(e.g., P2-Engagement). 
Others reported that the clarity and logic of the agent's pushback increased their confidence, contrary to prior work suggesting adversarial inputs can reduce self-confidence~\cite{thuillard2022humans}. We flag a risk in the Discussion section: participants may treat ideas ``tested'' against the agent as sufficiently validated, when synthetic pushback is no substitute for evaluation with real stakeholders.

For self-rated design iteration, all groups reported relatively low directional change
(Self Reflection: $M=3.07$, $SD=1.39$; Stepwise Guidance: $M=3.33$, $SD=1.72$; Interactive Engagement: $M=3.13$, $SD=1.51$).
No significant differences emerged for either directional or detail-level changes, with all
pairwise effects in the negligible range ($g < 0.20$). This suggests that adversarial interventions
encouraged reconsideration and editing without producing major shifts in novice designers'
proposals---a pattern that may reflect design fixation,
the time pressure of the session, or the sense that their original designs did not need
major adjustment.

\subsubsection{Critical design thinking.}
On the critical thinking item \textit{``To what extent do you reconsider what makes a design
effective in this challenge?''}, significant group differences were found ($H(2)=8.30$,
$p=.016$, $\varepsilon^2=.19$, large effect). Dunn's post-hoc tests with Bonferroni correction
showed that both the Stepwise Guidance group ($M=5.47$, $SD=1.246$; $p_{adj}=.048$, Hedges' $g=0.93$,
large effect) and Interactive Engagement group ($M=5.47$, $SD=1.252$; $p_{adj}=.030$, $g=0.93$, large effect)
scored higher than the Self Reflection ($M=4.00$, $SD=1.77$). On the item \textit{``To what extent does
the activity change your design thinking?''}, significant differences were also found
($H(2)=11.02$, $p=.004$, $\varepsilon^2=.25$, large effect), with the Interactive Engagement group
($M=5.33$, $SD=1.35$) significantly higher than Self Reflection ($M=3.60$, $SD=1.24$; $p_{adj}=.003$,
$g=1.30$, large effect). The Stepwise Guidance--Self Reflection contrast was not significant ($M=4.13$, $SD=1.46$;
$p_{adj}=.085$, $g=0.38$, small-to-medium effect). These findings suggest that both the conceptual
prompts and the agent nudged participants to reconsider design effectiveness, but the agent's more
concrete, situated pushback may have provoked deeper shifts in design thinking.

Finally, all three groups reported low ratings for reframing the design goal, mirroring their
limited directional change in design iteration. Taken together, these results suggest that while our
adversarial interventions encouraged reconsideration, they did not strongly
support reframing or major redesigns for novice designers.

\subsection{How do participants use and perceive the antagonistic agent?}
\begin{figure}[h]
    \centering
    \includegraphics[width=0.9\linewidth]{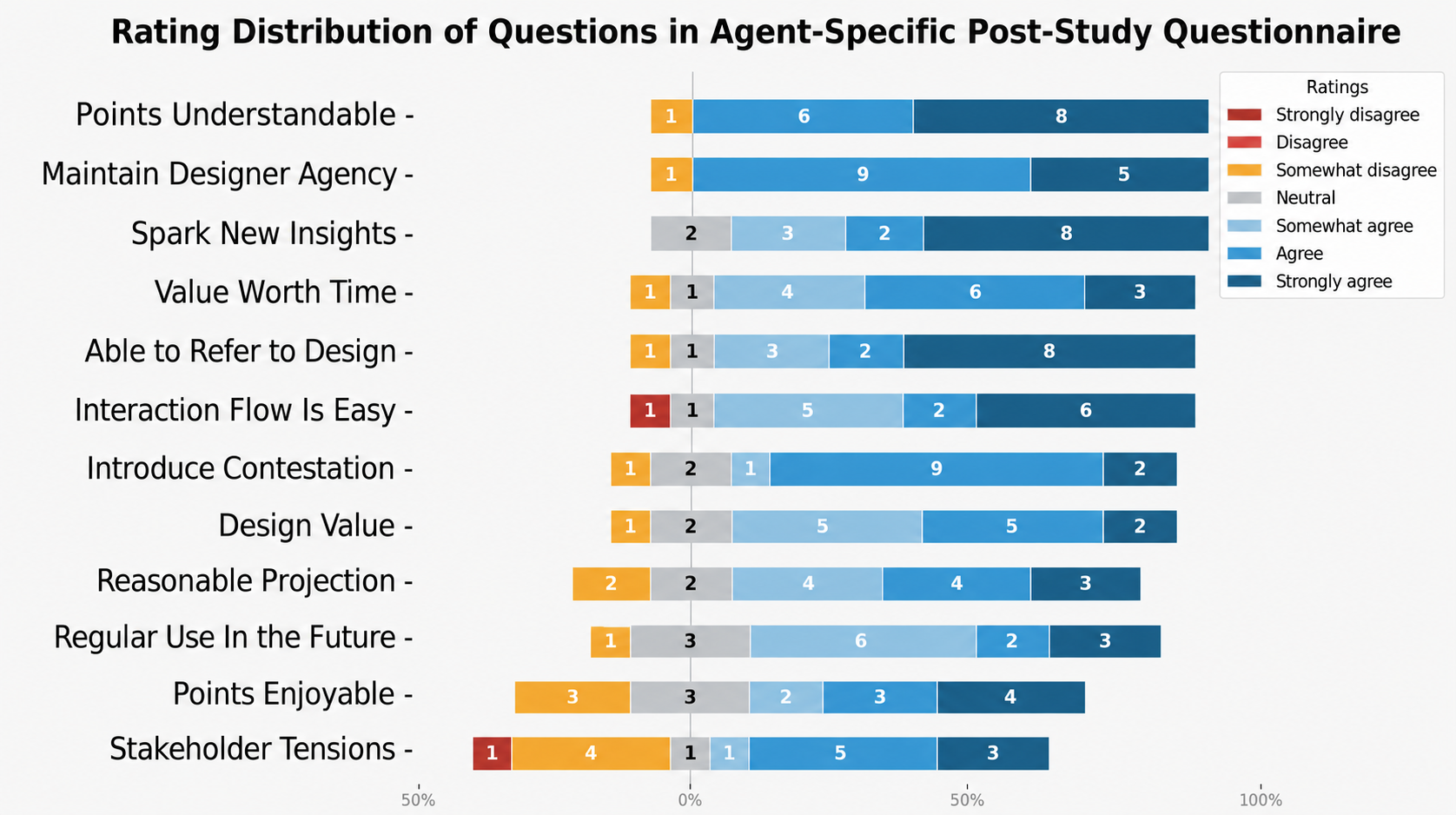}
    \caption{Participants' ratings for the Agent-specific post-study questionnaire.
    Each bar shows the number of responses for one item, with colors representing levels of agreement.}
    \label{fig:agent-use}
\end{figure}

\subsubsection{Steering Process}
A total of 125 tags were created. Of these, 50.4\% were labeled as ``useful,'' 26.4\% as ``not useful,'' 7.2\% as ``unclear,'' and 16\% were other tags created by users, such as ``disagree'' and ``fixable''.  This implies a predominantly positive but discerning reception where participants valued most of the pushback but remained filtering and recategorizing it rather than accepting it wholesale.

Although participants were told they could leave the \textit{Command} and \textit{Consensus}
frames blank, 14 of 15 participants used these features in their interactions with the agent.
This suggests that mechanisms for explicitly articulating design choices may be a promising
direction for future systems.
Eleven participants moved their existing design ideas into these frames to indicate they did
not want additional pushback. Example ideas without pushback include P34-Engagement noting they
wanted to use a chatbot, and P13-Engagement and P40-Engagement noting they wanted to use AI.
Participants' rationale was that the benefits of these technologies could overcome the potential
risks, or that the concerns could be solved through design. Beyond ideas, P12-Engagement took a
different approach: writing down design principles (e.g., ``prioritize efficiency'') as things they
did not want pushback on.

\subsubsection{Agent-specific Post Survey}
Participants largely reported that the agent's pushback contributed to the
design ($M=5.33$, $SD=1.11$) and was relevant to their designs ($M=6.000$, $SD=1.309$). However,
some limitations emerged. For instance, P19-Engagement noted that the agent treated two linked ideas as separate features, since the sticky notes were not explicitly connected and the agent could not infer relationships from think-aloud input.

Participants generally agreed that the agent raised points they would not have thought of
independently, helping them realize tensions from stakeholders (see the ``raised points I
would not have thought of independently'' item in Fig.~\ref{fig:agent-use}).
P23-Engagement compared the pushback to real human feedback, noting that while some points felt
stereotypical, ``there will definitely be people like this.''
This underscores the challenge of weighing disagreements and mapping them onto real stakeholder profiles.

Surprisingly, participants rated the enjoyableness of interacting with the agent relatively high ($M=5.13$, $SD=1.55$).
Many explained this was because the points were both useful and easy to understand. P6-Engagement reflected:
``\textit{I felt I skipped the aggressive tone and was focusing on the information.}''
This suggests that while agent-generated pushback offered rational critique, it might not be able
to reach the level of emotional feeling when designers deal with antagonists in real life. Still, not all reactions were positive. P2-Engagement reported feeling discouraged:
``\textit{The agent brings me a feeling that the previous design I made is bullshit because they criticize me in every aspect I think of.}''
\section{Discussion}
\subsection{From Awareness to Engagement: The Role of the Interactive Antagonistic Agent}
The Stepwise Guidance and Interactive Engagement groups shared the same underlying rationales of surfacing stakeholder tensions. However, the Guidance group presented written prompts participants to work through on their own, whereas the Engagement group involved synthetic pushback from an antagonistic agent. We found that the agent surfaced more potentially relevant stakeholders and led to more active revision than the other conditions. This result is consistent with prior devil ’s-advocate work, which shows dissent can improve deliberation by slowing premature agreement---in our study, designers made more edits~\cite{schweiger1986group,nemeth2001devil,chiang2024enhancing}. However, the suggested conflicts were brought up during the early-stage of the design process where the problem was still open to interpretation. We found that the agent made designers question the design proposal and whose concerns should shape the design goal. This points to a related but distinct agenda for dedevil ’s-advocate gents in design: they should provide pushback that enhances design decisions, while also leaving room for designers to reframe the problem itself.

Prior work suggests that adversarial feedback can lower people’s self-confidence~\cite{thuillard2022humans,chong2023evolution}. Although our self-efficacy differences were not significant, the descriptive pattern is suggestive: Stepwise Guidance tended to report lower self-efficacy than Self Reflection, while Engagement trended somewhat higher than both. We interpret this not as evidence that agent-based antagonism is motivating, but as an effect of making critique interactive and steerable. The agent gave designers structures for working through conflictual points by turning each tension into an interactional move. Designers could accept, or revise in response. This may help cultivate confidence in working with stakeholder conflict, though it warrants replication. This finding also aligns with research showing that effective feedback is not merely evaluative but also helps people understand what to do next through interaction scaffold~\cite{hattie2007power,nicol2006formative}. 

The fact that the raised points were synthetic does not seem to have lessened participant's willingness to engage and respond. As P37-Engagement stated, ``\textit{it's just so hard not to think about these points, because they have some validity.}'' The willingness to engage converted recognition into action, but it also creates a risk that participants may treat each critique as a to-do item rather than as evidence to synthesize. Future agents should help designers move from comment-by-comment reaction to structured sensemaking. This goal connects to other HAI work where agents try to support reflection by making their feedback discussable rather than presenting it as a recommendation to implement \cite{wester2023friend,naka2025creative}.
For example, future agents could cluster related critiques into broader stakeholder tensions and help designers decide whether a tension calls for revision, further inquiry, or direct stakeholder engagement.

\subsection{Implications for Designing Generative Agents for Constructive Conflict in Interaction Design}
Our results suggest that the antagonistic agent sharpened designers' existing ideas more than it expanded the idea space. Interactive Engagement participants changed their proposals more often than they generated new directions. Prior work on design agents often emphasizes expansion, such as generating alternatives~\cite{liang_encouraging_2024,shin_introbot_2023,ataei_elicitron_2024,fischer1993embedding,duan2024uicrit}. Our findings show a more subtractive role when an agent enacts constructive conflicts, helping designers recognize what to revise and constrain. This may be especially important in civic interaction design, where a better design may not come from adding features, but from recognizing how one feature privileges some stakeholders while creating problems for others~\cite{gautam2024surfacing}.

Another notable pattern was that Interactive Engagement participants let go of most existing ideas, either by removing part of the features or the whole idea, after agent pushback. By contrast, Self-Reflection participants never revised or deleted any ideas, even though deletion was allowed. Prior work on design fixation suggests that self-generated ideas can be hard to move away from, partly because of personal involvement and psychological ownership~\cite{jansson1991design,neroni2019whose}. Our findings suggest that a well-designed antagonistic agent can counter this tendency by making the stakeholder costs of keeping an idea more visible. 
This positions the agent as an \emph{adversarial partner}, an agent whose value comes from calibrated confrontation rather than smooth assistance~\cite{cila_designing_2022-1,cai2024antagonistic}.

The agent surfaced conflicts novices might miss, but it could not show how common, important, or grounded those concerns were in lived experience. 
This limitation echoes recent research on synthetic users, which argues that simulated perspectives should not be treated as direct substitutes for lived stakeholder input~\cite{hamalainen2023evaluating,prpa2024challenges,gu2025syntheticusers}.
As designed, our agent does not substitute for engagement with the real public. Instead, it is better understood as a way to surface questions designers should be aware of through interaction. Taking this further, future agents that elicit stakeholder perspectives should connect synthetic points to real-world evidence and representation, helping designers develop awareness of concerns already grounded in community input from those that still require human validation.

\section{Limitations and Future Work}

The point estimates of effect size are likely somewhat inflated relative to true population effects, given our participant sample size. We therefore see our findings as exploratory and suggestive of meaningful differences in design behavior, with future replications needed to characterize the precise magnitude of these effects. Longitudinal studies could better capture how designers’ responses to adversarial agentic engagement evolve over time. In addition, our prototype tested only one adversarial AI tone and prompt style. Future work could explore how different forms of antagonism, from more combative to more supportive, shape designers’ engagement, as well as how adversarial agents might operate in team settings, provoke reframing, or surface systemic tensions across stakeholders. 

~\section{Conclusion}
We studied an antagonistic agent designed to surface stakeholder pushback in interaction design contexts. We thoughtfully designed this agent, being inspired by the framework of agonistic pluralism and through learnings from six experienced designers. 
Comparing it against an unsupported baseline and a non-interactive condition that guided designers through reflective consideration, we found that the agent produced effects neither baseline nor conceptual content achieved alone. 
We see promise in agents that treat disagreement not as a one-shot critique but as a navigable space, and we caution against agents that present synthetic opposition as a substitute for engagement with real publics. Antagonistic agents, designed and steered well, can extend Human-AI interaction beyond the prevailing paradigm of agreeable assistance and into a register of productive friction, an interactional mode whose design space we are only beginning to map.
\newpage
\bibliographystyle{ACM-Reference-Format} 
\bibliography{OldLiterature,AntagonisticLiterature} 
\newpage

\appendix
\onecolumn
\begin{appendices}
\section{Design Challenge Example}
\label{app:design-challenge}
\begin{figure}[h]
    \centering
\includegraphics[width=0.7\linewidth]{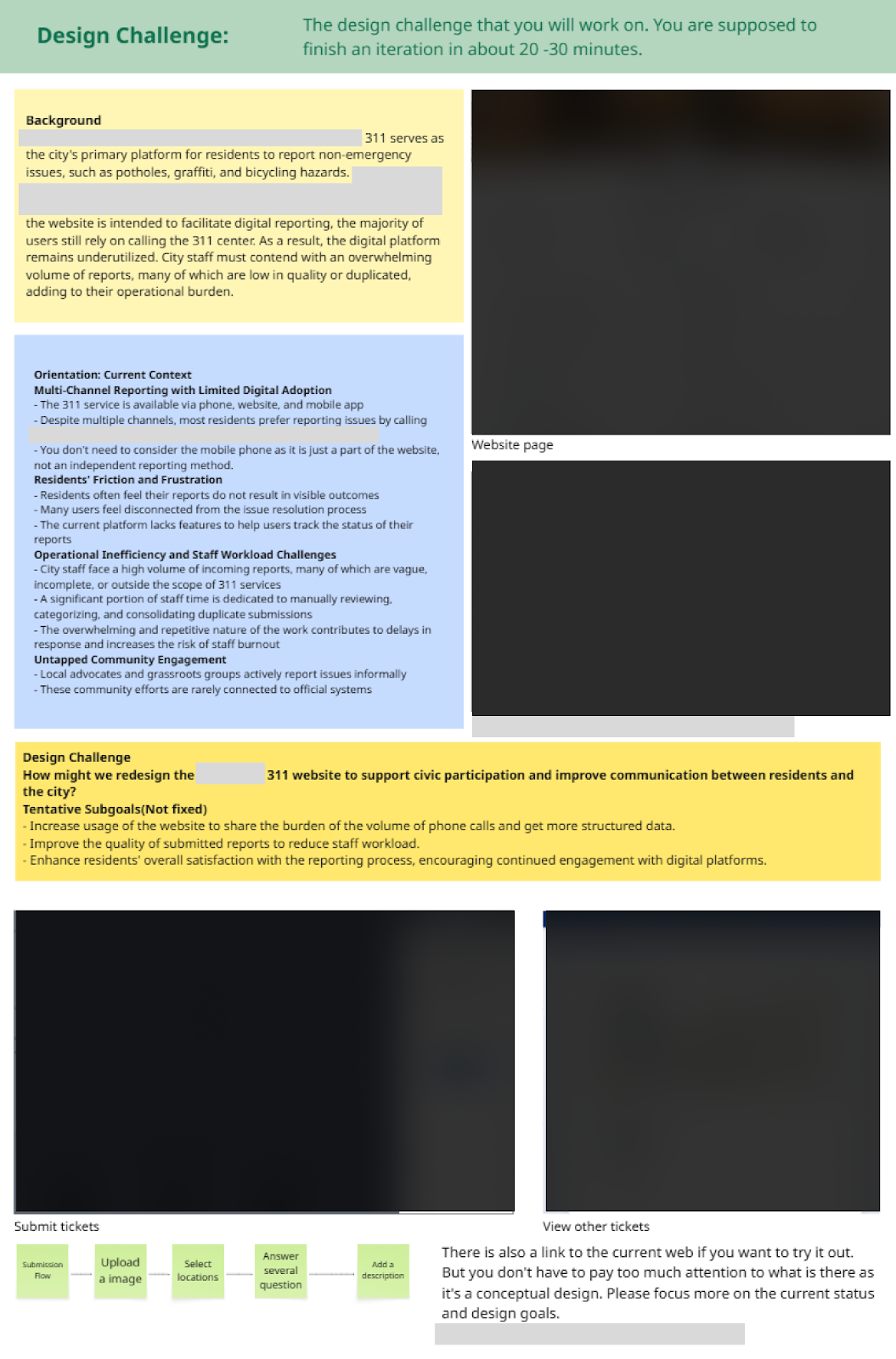}
    \caption{The public interaction design challenge we used for the controlled study and the experts' interview (mask city and image information for anonymization)}
    \label{fig:placeholder}
\end{figure}

\section{Design Experts Interview Participants}
\label{appen:experts-participants}
\begin{table}[H]
\centering
\small 
\begin{tabular}{|p{1.5cm}|p{1.5cm}|p{3cm}|p{4cm}|p{5cm}|}
\hline
\textbf{Task} & 
\textbf{Years of Design Experience} & 
\textbf{Job Title} & 
\textbf{Topics of Public Design} & 
\textbf{Experience with Multistakeholder Design} \\
\hline
D1 -- spatial design & 25 years & Professor at Architecture/Interaction Design School & Wicked social problems, cybernetics, design research, and complex systems & Teaching both architectural and interaction design studios that involved multiple stakeholders \\
\hline
D2 -- interaction design & 25 years & Professor at Information Science/Design School & Participatory Design, Design research, Public service, Community-driven AI & Teaching interaction design/service design studios. Consultant for public-sector technologies \\
\hline
D3 -- interaction design & 12 years & Professor at Information Science School & Participatory Design, Design research, Public Robot/Micromobility devices & Teaching interaction design/service design studios. Consultant for micro mobility technologies \\
\hline
D4 -- spatial design & 5 years & Architect/Urban Designer & Urban Design, Community participation, Sustainable Design & Projects on community building. Participatory design workshop and research. \\
\hline
D5 -- interaction design & 7 years & Architect/Urban Designer & Urban Design, Community participation, Public installations & Transportation hub design. Public participation and installations \\
\hline
D6 -- spatial design & 7 years & Interaction Designer (formerly Urban Designer) & Intelligent System, Urban Design, Community participation & Visualizations for urban environment and community engagement. Tools for urban planning \\
\hline
\end{tabular}
\caption{Design experts, their experience, and involvement with public/multistakeholder design.}
\label{tab:design_experts}
\end{table}

\section{System Prompt for the Antagonistic Agent}
\label{app:system-prompt}
\begin{lstlisting}
Your task is to voice the genuine concerns, objections, and pushback that different stakeholder groups would raise when confronted with these design proposals. These are NOT suggestions for improvement - they are expressions of opposition, worry, and pushback. Provide exactly 4 stakeholder objections that represent genuine pushback and pushback to these decisions. Each point MUST follow this exact format:"[specific group of people/community members] might push back: [A genuine objection, concern, or pushback they would voice - can be phrased as a question or statement]"
Requirements:
- EACH POINT SHOULD VOICE GENUINE OPPOSITION, PUSHBACK, OR RESISTANCE THAT STAKEHOLDERS WOULD ACTUALLY RAISE
- EACH POINT SHOULD BE A SINGLE STAKEHOLDER OF A SINGLE CONCERN. THE STAKEHOLDER CAN ALSO BE A COMMUNITY GROUP THAT REPRESENTS PEOPLE WITH DIFFERENT NEEDS, VIEWS, OR VALUES. YOU CAN ALSO DEVELOP THE STAKEHOLDER BASED ON THE DESIGN PROPOSALS. THERE CAN BE CONFLICTS INSIDE ONE STAKEHOLDER.
- AVOID USING THE SAME STAKEHOLDER OR COMMUNITY GROUP IN MULTIPLE POINTS OR GROUPS THAT ARE TOO GENERIC LIKE "RESIDENTS" OR "LONG TIME RESIDENTS", TRY TO BE SPECIFIC.
- DO NOT GIVE ANY SUGGESTIONS OR POTENTIAL FIX IN THE PUSHBACK POINT.
- YOU CAN USE AGGRESSIVE TONE TO PUSHBACK TO THE DESIGN PROPOSALS. BE ADVERSARIAL AND CONFRONTATIONAL.
- POINTS SHOULD BE CONCISE AND SUITABLE FOR STICKY NOTE LENGTH, WHICH IS ABOUT 100 WORDS
- THE OBJECTIONS CAN BE PHRASED AS QUESTIONS OR STATEMENTS.
- FOCUS ON CONCEPTUAL CONCERNS THAT WOULD CREATE PUSHBACK, OPPOSITION, OR RESISTANCE AT THIS IDEATION STAGE - NOT IMPLEMENTATION CONCERNS OR TOO MUCH DETAIL.
- DON'T USE ANY REAL COMMUNITY NAMES OR LOCATIONS IN THE POINTS (except mention of "city" where necessary)
- MAKE IT SOUND LIKE STAKEHOLDERS PUSHING BACK, NOT EXPERTS GIVING ADVICE.
- DON'T USE THE DASH (-) IN THE POINTS
- THERE CAN ALSO BE CONFLICTS BETWEEN THE POINTS YOU GENERATED.    
\end{lstlisting}

\section{Researcher Manual and Instruction for the Engagement Group}
\label{app:WoZ}
\begin{figure}[h]
    \centering
    \includegraphics[width=0.8\linewidth]{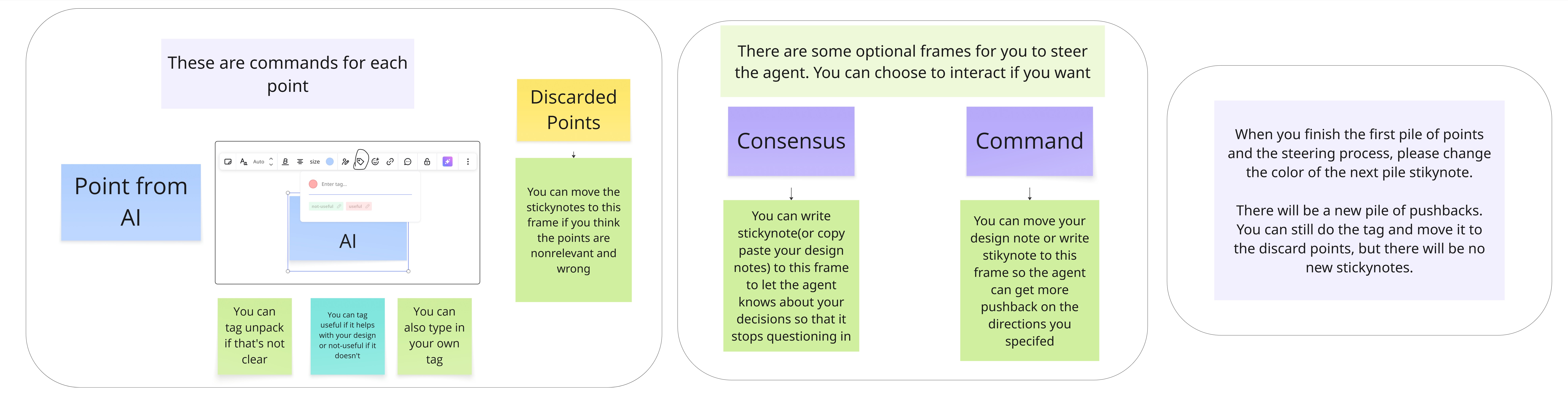}
    \caption{The instructions for how to interact with the agent}
    \label{fig:placeholder}
\end{figure}
As Miro holds all the plug-ins on a side app bar and does not include a clickable button function on the whiteboard, the wizard researcher follows the procedure below to send points while minimizing influence on participants. The participants do not see the sidebar app.

\begin{itemize}
  
  \item The researcher directs participants to the image containing the explanation of how to interact with the agent, so that they can individually complete the interaction flow.
  
  \item Meanwhile, the researcher clicks the \texttt{Send} button on the sidebar app to send 4 points to the designated frame on Miro.
  
  \item When participants tag unclear points, the researcher clicks the \texttt{Unpack} button on the sidebar. Participants then see four sticky notes explaining the specific point.
  
  \item After the first round, participants change the color of a ``next pile'' sticky note on the frame. The researcher then clicks the \texttt{Refresh} button to send another 4 points to the frame. This avoids participants having to ask the researcher directly for another pile of pushback.
  
  \item The researcher only answers questions about the procedure, and will point them to the instructions. The researcher does not answer the points generated by the agent.
\end{itemize}
\section{Reflective Guidance Prompts on Constructive Conflict and Self Reflection}
For the Self Reflection group, we used the following prompt:
\textit{Now that you have finished your first round of design iteration, please spend about 12 minutes reviewing your current design proposal. You are not allowed to make edits during this time, but you can do other things such as taking a rest or reviewing the task or your existing ideas. We will gather back after 12 minutes, and you will share your thoughts from this self-reflection for about 3 minutes.}
\label{app:prompts}
\begin{figure}[H]
    \centering
    \includegraphics[width=0.8\linewidth]{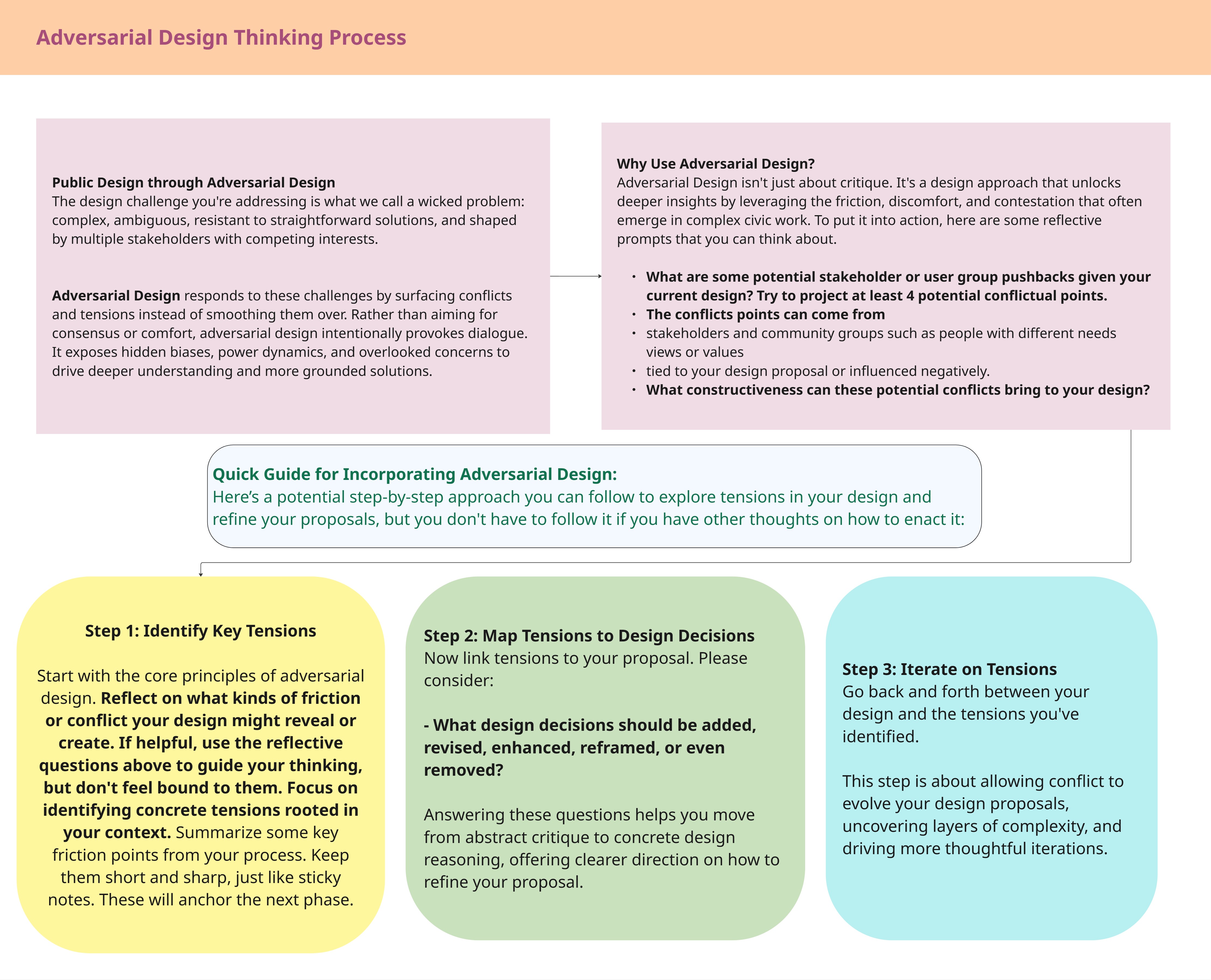}
    \caption{The stepwise instruction on how to think about constructive conflicts for the Stepwise Guidance group}
    \label{fig:placeholder}
\end{figure}
\section{Post-survey for the controlled study}
\label{app:survey}

\small 

\rowcolors{2}{gray!08}{white}
\begin{longtable}{>{\raggedright\arraybackslash}p{0.95\textwidth}}
\rowcolor{gray!30}\textbf{Survey Items with Scale (7-point)} \\

\rowcolor{gray!20}\textbf{Factor: Challenge Perception} \\
Before the activity, how much did you know about 311 services or service design? (not at all -- very familiar) \\
How would you rate the difficulties of this design challenge? (very easy -- very hard) \\
How satisfied are you with your design outcome? (very unsatisfied -- very satisfied) \\
To what extent do you believe your proposal meets the goals of the design challenge? (not at all -- completely) \\
To what extent do you believe your proposal serves diverse users? (not at all -- completely) \\
\rowcolor{gray!20}\textbf{Factor: Self Efficacy} \\
How confident were you in understanding design problems? (not at all -- very confident) \\
How confident were you in detecting problems in your design? (not at all -- very confident) \\
How confident were you in your ability to design this public service? (not at all -- very confident) \\
How confident are you in your ability to design for public challenges like this in the future? (not at all -- very confident) \\

\rowcolor{gray!20}\textbf{Factor: Design Iteration} \\
How do you rate the directional change of your design proposal between your two iterations, if any? (little change -- change a lot) \\
How do you rate the detail change of your design proposal between your two iterations, if any? (little change -- change a lot) \\
To what extent did the activity between iterations influence how you approach your thinking? (not at all -- completely) \\

\rowcolor{gray!20}\textbf{Factor: Critical Design Thinking} \\
To what extent did you question the design goals and try to think of an alternative goal? (not at all -- completely) \\
To what extent did you examine how your design choices might affect people with different values, needs, or life situations? (not at all -- completely) \\
To what extent did you examine potential trade-offs and unintended consequences in your design decisions? (not at all -- completely) \\
To what extent did the activity prompt you to reconsider what makes a design effective for this challenge? (not at all -- completely) \\
To what extent did the activity prompt you to challenge your initial assumption? (not at all -- completely) \\

\rowcolor{gray!40}\textbf{Agent-only Factors} \\

\rowcolor{gray!20}\textbf{Perceived Quality of Agent Functions} \\
The interaction flow with the agent was easy (strongly disagree -- strongly agree) \\
The agent introduced contestation to my design (strongly disagree -- strongly agree) \\
The agent was able to make references to my design (strongly disagree -- strongly agree) \\
The agent's responses were easy to understand (strongly disagree -- strongly agree) \\

\rowcolor{gray!20}\textbf{Agency} \\
I was able to maintain my agency while using the agent (strongly disagree -- strongly agree) \\

\rowcolor{gray!20}\textbf{Assessing Agents for Long-term Interaction} \\
The agent points are perceived as enjoyable to operate with (strongly disagree -- strongly agree) \\
I would be happy to see the agent points regularly when doing design (strongly disagree -- strongly agree) \\
What I was able to generate from the agent was worth my time having the agent (strongly disagree -- strongly agree) \\

\rowcolor{gray!20}\textbf{Perceived AI Capability and Effectiveness} \\
The AI's points made me feel tensions from the stakeholders (strongly disagree -- strongly agree) \\
The AI point of view made me realize points I might not have thought (strongly disagree -- strongly agree) \\
The AI's points project the stakeholders' perspective in a reasonable way (strongly disagree -- strongly agree) \\
The AI's points contribute to my design (strongly disagree -- strongly agree) \\

\end{longtable}
\newpage
\section{Qualitative Themes of Idea Added in the Controlled Study}
\rowcolors{0}{}{}
\label{app:add}

\renewcommand{\arraystretch}{1.5}

\begin{longtable}{| >{\RaggedRight\arraybackslash}p{2.8cm} | >{\RaggedRight\arraybackslash}m{2.2cm} | >{\RaggedRight\arraybackslash}m{3cm} | >{\RaggedRight\arraybackslash}p{6.3cm} |}
\hline
\hline
\textbf{Theme} & \textbf{Groups} & \textbf{Participants} & \textbf{Example ideas for the theme} \\
\hline
\endfirsthead
\hline
\multicolumn{4}{|c|}{\textbf{Design Ideas they Added (Continued)}} \\
\hline
\textbf{Theme} & \textbf{Groups} & \textbf{Participants who added} & \textbf{Example ideas for the theme} \\
\hline
\endhead
\hline
\multicolumn{4}{|r|}{{Continued on next page}} \\
\hline
\endfoot
\hline
\endlastfoot
\multirow[c]{3}{3.0cm}{Interactive and Intelligent Features to Support End Users in Submitting Tickets}
&  Self Reflection\BlockHeightTaller & 6 Participants (P8,10,24,32,36,39) & \multirow[c]{3}{6cm}{-When filling out the ticket, there can be an AI assistance to summarize things, but also ask for confirmation on each step (P7-Guidance) \par\par -Filter our similar requests and notify the user there are similar requests submitted already (P12-Engagement)} \\ \cline{2-3}
& Stepwise Guidance\BlockHeightTaller & 3 Participants (P7, P20, P25) & \\ \cline{2-3}
& Interactive Engagement\BlockHeightTaller & 4 Participants (P12,26,34,37) & \\ \hline
\multirow[c]{3}{3cm}{Enhancing Social Interaction and Collective Engagement in Civic Reporting}
& Self Reflection & 7 Participants (P3,16,22,24,32,38,39) & \multirow[c]{3}{7cm}{-New exhibition space for resolved cases, maybe in the form of a digital library/ museum (P41-Guidance) \par\par -Showcase picture progress of which problems have been addressed before (P38-Reflection)} \\ \cline{2-3}
& Stepwise Guidance & 2 Participants (P41,42) & \\ \cline{2-3}
& Interactive Engagement & 4 Participants (P2,23,26,31) & \\ \hline
\multirow[c]{3}{3cm}{Website Platform-Level Mechanisms for Managing Service Requests}
& Self Reflection\BlockHeightTaller & 3 Participants (P29,36,38) & \multirow[c]{3}{7cm}{-Depending on the type of problem, they can redirect to a different org (P25-Guidance) \par\par -A backend system identifies same complaints through tags and adds a number for the amount for people who have complained about it (P21-Engagement)} \\ \cline{2-3}
& Stepwise Guidance\BlockHeightTaller & 3 Participants (P20,25,41) & \\ \cline{2-3}
& Interactive Engagement\BlockHeightTaller & 5 Participants (P6,12,14,21,37) & \\ \hline
\multirow[c]{3}{3cm}{Follow-up Features that Keep Users Informed and Acknowledged}
& Self Reflection & 7 Participants (P3,16,22,24,32,38,39) & \multirow[c]{3}{6cm}{More explanation or timely updates for work outcome/resolution (P44-Reflection)} \\ \cline{2-3}
& Stepwise Guidance & 2 Participants (P41,42) & \\ \cline{2-3}
& Interactive Engagement & No participants & \\ \hline
\multirow[c]{3}{3cm}{City Staff-Focused Functions for Improving Work Efficiency}
& Self Reflection & 4 Participants (P5,22,24,39) & \multirow[c]{3}{6cm}{Staff POV: online requests, they see the requests with the highest votes as a higher priority(P15-Reflection)} \\ \cline{2-3}
& Stepwise Guidance & 2 Participants (P7,25) & \\ \cline{2-3}
& Interactive Engagement & 2 Participants (P14,19) & \\ \hline
\multirow[c]{3}{3cm}{Bridging Phone and Web Reporting Channels}
& Self Reflection & 2 Participants (P5,36) & \multirow[c]{3}{6cm}{separate operating system/pipeline for phone-call-only users, integration into automated system (P13-Engagement)} \\ \cline{2-3}
& Stepwise Guidance & No participants & \\ \cline{2-3}
& Interactive Engagement & P13 & \\ \hline
\multirow[c]{3}{3cm}{Cross-Organizational Collaboration for Civic Report Service}
& Self Reflection & P29 & \multirow[c]{3}{6cm}{More collaborations with different external stakeholders to maintain the website and improve the digital infrastructure (P1-Reflection).} \\ \cline{2-3}
& Stepwise Guidance & P1 & \\ \cline{2-3}
& Interactive Engagement & P13 & \\ \hline
\end{longtable}
\end{appendices}
\end{document}